# Stability of the $N = 2$ bubble states in Corbino geometry under a tilted field

Pengjie Wang[1], Yijia Wu[1], Hua Chen[1], Ruoxi Zhang[1], L. N. Pfeiffer[2], K. W. West[2], Rui-Rui Du[1,3], X. C. Xie[1,3], and Xi Lin[1,3,†]

1 International Center for Quantum Materials, Peking University, Beijing 100871, China
2 Department of Electrical Engineering, Princeton University, Princeton, New Jersey 08544, USA
3 Collaborative Innovation Center of Quantum Matter, Beijing 100871, China

[†]xilin@pku.edu.cn

**ABSTRACT**

We measured the activation energy gap of the $N = 2$ bubble states using tilted-field technique in two GaAs/AlGaAs samples with Corbino geometry, and observed monotonically decreasing energy gaps with the increasing total magnetic field. The observed tilt angle dependence is discussed based on magneto-orbital coupling and finite-size effect. We also found a bubble state at an uncommon filling factor of 4.67, with an unanticipated energy gap of up to 5.35 K and opposite in-plane field dependence.

**Introduction**

The electronic crystals are of the spontaneous superstructure formed by electrons in appropriate conditions [1-4]. In recent years, the nematic phases in different kinds of materials, such as ruthenates [5], high-$T_C$ superconductors [6,7], and heavy fermion materials [8], have been observed and drawn continuous attention. In contrast to the anisotropic nematic phases, there are isotropic electron solids such as Wigner crystals [9], which may occur at ultra-low temperature and high magnetic field in two-dimensional electron gas (2DEG) system.

In an ultra-high mobility 2DEG based on GaAs/AlGaAs heterostructures, both anisotropic and isotropic phases have been observed at the high Landau level ($N > 1$) [3,10-19]. The anisotropic phase is named as stripe phase, in which the longitudinal resistance along different orientation is strikingly different: one manifests itself as a peak, while the other a dip. These interesting behaviors have been further observed at the second Landau level ($N = 1$) under a tilted magnetic field [20] and hydrostatic pressure [21]. The isotropic phase, in which the Hall resistance reentrants into the nearest integer quantum Hall (IQH) state, is called reentrant integer quantum Hall

(RIQH) state. A RIQH state is believed to be a bubble phase, in analogy to Wigner crystals with multiple electrons per site [10,11].

RIQH states are typically studied by transport measurements with continuous interest for years [15,22-32]. Bubble phases have been also confirmed by the resonance in the real part of the conductivity under microwaves [23], and have been measured by surface acoustic waves [33]. Most of the existing experiments focus on the non-linear dynamics and phase transition between electron liquid and solids. Experiments probing the influence of the in-plane magnetic field on the bubble state itself are rare, with only one exception on the second ($N = 1$) Landau level [26].

In this paper, we studied the stability of the bubble states in the third Landau level ($N = 2$) using tilted-field technique. The bubble states in the third Landau level are the strongest and most convenient to investigate the influence of an in-plane magnetic field. Corbino disc samples were used in order to exclude the effect of Hall electrical field and to measure the conductivity directly. We observed non-linear decreasing energy gaps at bubble states with increasing total magnetic field. This decreasing behavior is related to magneto-orbital coupling together with finite-size effect of the quantum well. We also observed an unexpected bubble state at filling factor 4.67 with opposite in-plane-field dependence, which has an energy gap of above 4 K, in contrast to common energy gap of ~ 1 K for a bubble state.

**Samples and Experiments**

Our experiments were conducted on two modulation doped GaAs/AlGaAs quantum wells. The densities of Sample A and B are $4.2\times10^{11}$ cm$^{-2}$ and $2.8\times10^{11}$ cm$^{-2}$ with the mobilities $1.2\times10^7$ cm$^2$ V$^{-1}$ s$^{-1}$ and $2.8\times10^7$ cm$^2$ V$^{-1}$ s$^{-1}$, respectively. Both samples are patterned as Corbino geometries. The quantum well widths of Sample A and B are 25 nm and 28 nm. Sample A (B) has an inner diameter of 1.8 mm (1.4 mm) and an outer diameter of 2.0 mm (1.6 mm).

The samples were cooled down in a Leiden CF-CS81-600 dilution refrigerator equipped with a 9 Tesla magnet and a measurement probe. An attocube piezo-driven rotator was adapted to provide continuously in-situ sample rotation. The tilted angle was determined by the Shubnikov–de Haas oscillation of the Corbino samples and checked independently by a van der Pauw sample in the same cool down. The base temperature at the sample holder was 16 mK with a base electron temperature 22 mK. Illumination with a red light-emitting diode (LED) was applied at 4.5 K with 15 µA for 1 hour before measurements were taken. The measurements were based on a standard lock-in technique at 17 Hz. The maximum ac voltage drop across Corbino samples was 50 µV.

The differential conductivity of Sample A at a series of temperature with the magnetic field perpendicular to the 2DES (tilted angle $\theta = 0°$) is provided in Fig. 1(a). IQH states, fractional quantum Hall (FQH) states and RIQH states all result in a zero conductance for a Corbino disc. The longitudinal resistance and the Hall resistance in a van der Pauw sample (easy direction, from the same wafer of Sample A) is shown in Fig. 1(b), which demonstrates corresponding IQH plateaus at fractional filling factors and confirms the existence of RIQH states. For clarity, we label the IQH state with filling factor 4 as IQHE4, and the RIQH state that reentrants to the neighboring IQHE4 as RIQHE4. The insets of both figures are the schematics for the measurement setups.

By tilting the sample at a given temperature, the effect of the in-plane magnetic field can be explored qualitatively. In Fig. 2, the differential conductivity of sample A is plotted as a function of vertical magnetic field near RIQHE5 at 90 mK. The minimum of the differential conductivity represents the stability of the bubble state. The higher the minimum, the weaker the bubble state is. Therefore, the RIQHE5 bubble state shown here is suppressed by the in-plane magnetic field.

In order to quantitatively investigate the effect of the in-plane magnetic field, we measured the activation energy gaps of RIQHE4 and RIQHE5 at a series of angles. The energy gap, $\Delta$, can be derived from the temperature dependence of the corresponding conductivity minimum in the thermally activated regime where $\sigma_L \propto \exp(-\Delta/k_B T)$ [15]. In Fig. 3, the gaps of RIQHE4 (Sample A), RIQHE5 (Sample A) and RIQHE4 (Sample B) are plotted as a function of total magnetic fields, $B_{total}$, with a set of temperature-dependent curves shown in an Arrhenius plot inset in each panel. The gaps decrease with the tilted angle. Surprisingly, the energy gaps are not linear with the total magnetic field ($B_{total}$), which is unexpected if only the Zeeman effect plays a role.

In Fig. 4(a), we observed a contrasting increasing energy gap versus total magnetic field in RIQHE5 of Sample B. Such an increasing gap has never been reported before. The filling factor of RIQHE5 (Sample B) is 4.67, away from the typical value of 4.75, and the energy gap itself is up to above 4 K, significantly larger than the typical ~ 1 K for bubble states in the third Landau level. Shockingly, the RIQHE5 of the van der Pauw sample from the same wafer is normal, for all of the filling factor, the energy gap magnitude and the magnetic field dependence (Fig. 4(b)). The difference between the Corbino sample and the van der Pauw sample for the filling factor (Fig. 4(c)) and energy gap behavior (Fig. 4(a) and Fig. 4(b)) may come from the inhomogeneity of the wafer, which causes at least the electron density difference between two samples.

**Decreasing Energy Gaps of RIQH States**

**a) Spin Polarization**

In an ideal 2DEG system, the quantum well width is approximated to be zero. Applying an in-plane magnetic field by rotating the sample causes the coupling of the electron spin with the total magnetic field through the Zeeman energy. The Zeeman energy can be derived from $E_Z = g^* \mu_B B_{total} S$, where $g^*$ is the effective electron $g$-factor, $\mu_B$ is the Bohr magneton, $B_{total}$ is the total magnetic fields and $S$ is the electron spin. With this, an energy gap induced by spin coupling should yield a universal slope in a plot of activation gap versus $B_{total}$. Extracting the slopes of the energy gap from the RIQH states (Fig. 3), we obtained $g$-factor values of 0.84±0.18, 0.42±0.09, and 0.70±0.10 for RIQHE4 (Sample A), RIQHE5 (Sample A) and RIQHE4 (Sample B), respectively. One of them fits the commonly accepted effective $g$-factor in GaAs, i.e. -0.44.

In this way, the direct interpretation for our observed decreasing energy gap (Fig. 3) seems to be that RIQHE4 (Sample A), RIQHE5 (Sample A) and RIQHE4 (Sample B) are spin-unpolarized. Normally, the filling factors are 4+1/4 and 4+3/4, and the bubble states are expected as the two-electron bubble based on calculations [34-36], so it is reasonable to speculate that two-electrons-per-bubble leads to a spin unpolarized state. A spin unpolarized state could be destroyed at higher tilt, due to the Zeeman splitting under a high magnetic field [37]. However, in our data, when the in-plane magnetic field further increases, the energy gaps deviate from the decreasing trend and saturate with a total magnetic field at a few hundred millikelvins. These saturations may indicate a fully spin polarized bubble state. Such a discrepancy suggests that the tilted field technique itself should be insufficient to determine the spin-polarization [37]. Other mechanisms, like energy level crossing and finite-size effect of quantum well width, could also induce a decreasing energy gap. A more direct measurement technique, such as resistively detected nuclear magnetic resonance [38], is required to determine the spin configuration.

**b) Energy Level Crossing**

The energy gaps aren't linear over the entire range as a function of total magnetic field in Fig. 3, indicating that the Zeeman splitting is not the only cause of gap weakening. A non-linear decreasing energy gap could occur when two energy levels get too close and we should discuss the role of the energy level crossing in our tilted measurement. In our system, crossing between the opposite spin branch of the nearest Landau level is difficult to happen. For a 2DEG in GaAs/AlGaAs heterostructures, the effective mass of the carriers is too small for these two energy levels to cross over each other. For our samples, a total magnetic field larger than 200 T is required to cause the crossing, which is a value too high for steady field magnets.

It's also impossible for the energy level crossing to occur between the $N = 0$ Landau level of asymmetry subband (A0) and the $N = 2$ Landau level of symmetry subband (S2). The cyclotron energy, $\hbar\omega_c$, can be obtained by the vertical magnetic field of the filling factor. In sample A (B), they are 81.7 K (52.6 K) for RIQHE4 and 74.4 K (49.7 K) for RIQHE5. With estimations from an infinite quantum well with specified width, the energy gaps

$\Delta_{SAS}$ between the asymmetry ground subband (A0) and symmetry ground subband (S0) are 312.6 K for sample A and 249.2 K for sample B, which are larger than twice of the cyclotron energy. Therefore, in both samples, the A0 level should be higher than the S2, as shown in Fig. 4(d). In an early work on wide quantum well ($w$ = 55 nm) [39], stable $q/3$ FQH states are observed in high ($N > 1$) Landau level when Fermi level lies in A0 level. If the crossing between S2 level and A0 level does happen, as shown in Fig. 4(e), then we may observe the hint of $q/3$. We didn't see any evidence for FQH states in the third Landau level of both samples with a sufficient low temperature environment and high sample mobility.

**c) Magneto-orbital Coupling induced Finite-size Effect**

When the parallel magnetic length, $l_\parallel = \sqrt{\hbar/eB_\parallel}$, induced by the in-plane magnetic field, is comparable or less than the quantum well width, the magneto-orbital coupling between the in-plane magnetic field and transverse electron becomes possible [40]. In this circumstance, even a spin-polarized bubble state, which is believed to be stabilized with tilt, can be suppressed by increasing in-plane magnetic field $B_\parallel$, in addition to the Zeeman coupling.

In Fig. 5, we replotted the normalized energy gap values, $\Delta/\Delta_{B_\parallel=0}$, as a function of in-plane magnetic field. All the normalized gaps aligned well with each other. And decreasing trend for the normalized energy gaps under the in-plane magnetic field occurred, which suggests that the suppression from the finite-size effect was possible. Moreover, the decreasing trend is linear. Such a linear relation for the activated energy gap versus in-plane magnetic field, rather than the total magnetic field, have been also observed in a previous experiment on the 5/2 state under a tilted magnetic field [41], in which it was attributed to magneto-orbital effect in a finite-size quantum well [42].

From the discussion above, the most possible mechanism for such a decreasing gap is the finite-size effect of quantum well width through magneto-orbital coupling.

**Anomalous Increasing Energy Gap**

Although most of the RIQH states share universal behavior under tilt, we observed an anomalous energy gap increasing at RIQHE5 in sample B, as shown in Fig. 4(a). The filling factor of this state is centered at 4.67, which is close to 4+2/3, different from the usually observed 4+3/4. In a differential conductance measurement, both FQH states and RIQH states lead to an insulating behavior in Corbino geometry. However, FQH states and RIQH states can be distinguished by different non-linear behaviors mentioned in Ref. [30]. In our experiments, depinning behaviors similar to Ref. [30] are observed in all tilted angles we examined.

The energy gap without tilt for this state is around 4 K, while at large tilt it's 5.35 K. We also note that the gap increase at this RIQH state is not linear. At the low tilt regime, the increasing rate is about 4 times as large as the single particle Zeeman coupling alone (Fig. 4(a)), in which might suggest a non-trivial excitation spectrum. A spin texture similar to the skyrmion excitation, observed in previous $v = 1$ IQH [43] and $v = 1/3$ FQH states [44,45], might be involved. The formation of the skyrmions is characterized by the interplay between Zeeman energy and Coulomb energy, $E_C = e^2/4\pi\varepsilon\ell_0$, where $\varepsilon = \varepsilon_r\varepsilon_0$, and $\varepsilon_r \approx 13$ is the dielectric constant of GaAs and $\ell_0 = \sqrt{\hbar/eB_\perp}$ is the magnetic length without tilt [43,46]. In early experiments, the skyrmionic spin-reversal excitation are observed only to exist below the critical value of the ratio, $\eta = E_Z/E_C$. In this anomalous RIQH state, $\eta \sim 0.01 - 0.015$ is below $\eta_c = 0.022$ in $v = 1$ [43], and similar to $\eta_c = 0.01$ in $v = 1/3$ [44]. Therefore, it holds the possibility that such an anomalous energy enhancement may arise from the existence of skyrmions involving 4 reversed spins [43,44].

**Conclusion**

In summary, we observed monotonically decreasing energy gaps with increasing total magnetic field at RIQHE4 and RIQHE5. This effect could be understood by the magneto-orbital coupling induced finite-size effect of the quantum well width. A contrasting increasing energy gap at filling factor 4.67 is observed and it's interpreted as the formation of skyrmions consisting of a small number of spin-reversal excitations. The information of spin polarization of the RIQH states cannot be determined by tilting a magnetic field in our experiment. More sensitive techniques, such as resistively detected nuclear magnetic resonance, are desirable.

**Acknowledgement**

We are grateful to Koji Muraki, Xin Wan, and Kun Yang for helpful discussions. The work at Peking University was funded by NBRPC (Grant No. 2015CB921101) and NSFC (Grant No. 11674009, 11274020 and 11322435). The work at Princeton University was funded by the Gordon and Betty Moore Foundation through the EPiQS initiative Grant GBMF4420, by the National Science Foundation MRSEC Grant DMR-1420541, and by the Keck Foundation.

**Figures**

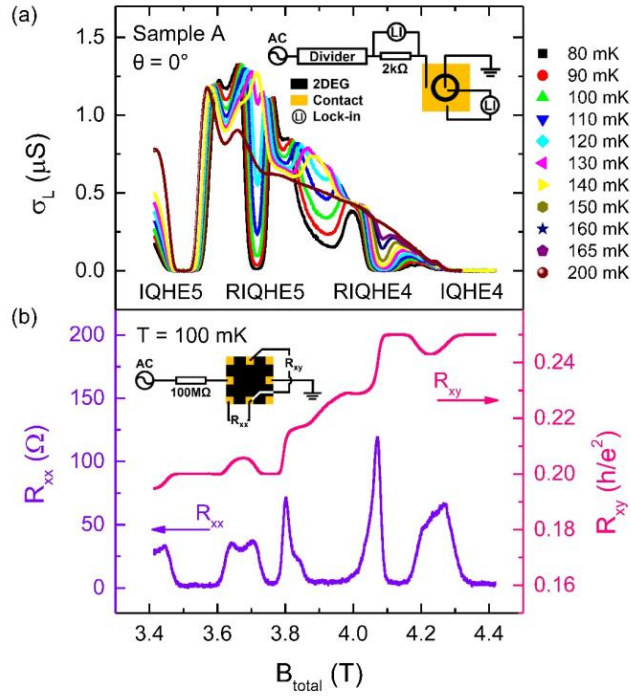

FIG. 1. (a) Longitudinal conductivity of Sample A versus total magnetic field at different temperatures without tilt. Inset: differential conductivity measurement setup for Corbino samples. (b) Longitudinal and Hall resistance of a van der Pauw sample (easy direction) made from the same wafer as Sample A. Inset: Hall measurement setup for van der Pauw samples.

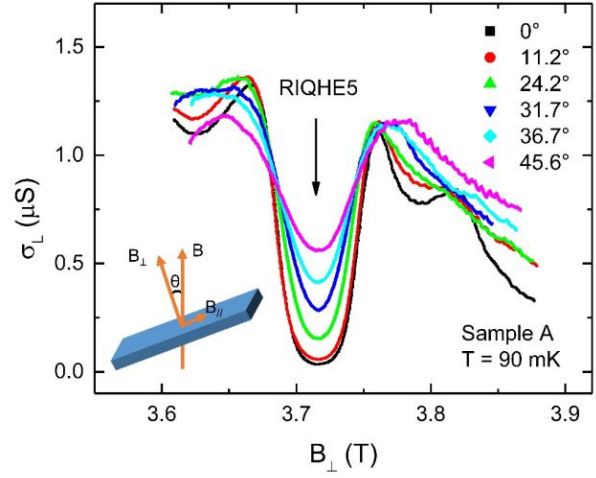

FIG. 2. Differential conductivity of Sample A versus vertical magnetic field under different tilted angles at 90 mK. Inset: a schematic for tilted magnetic field technique.

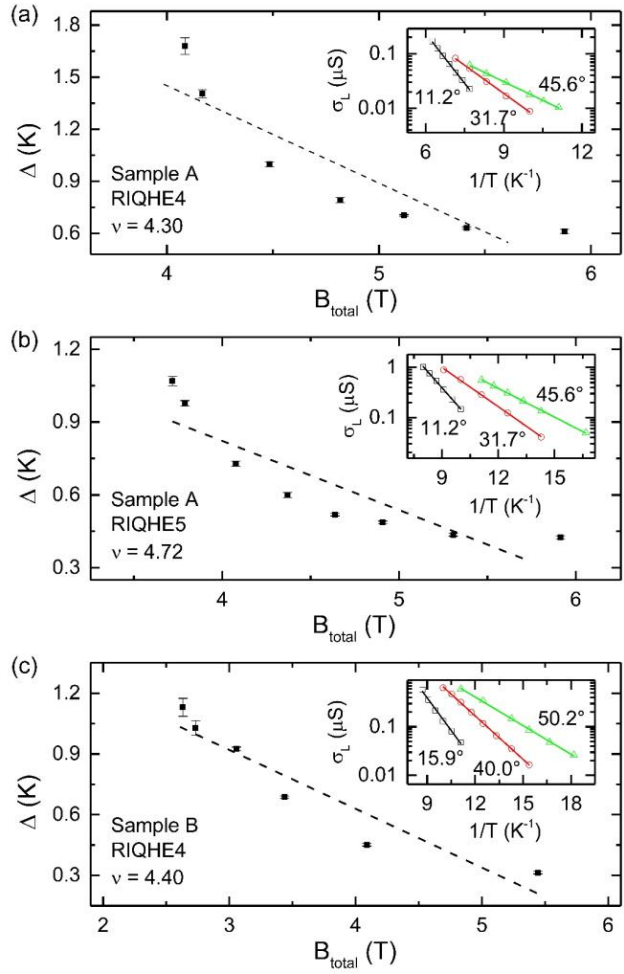

FIG. 3. Activation gaps versus total magnetic field (corresponding Arrhenius plots are shown inset). (a) and (b) are the energy gap variation of RIQHE4 and RIQHE5 taken from Sample A, (c) is the energy gap variation of RIQHE4 taken from Sample B. Linear fits (dashed lines) are used to estimate the *g*-factor.

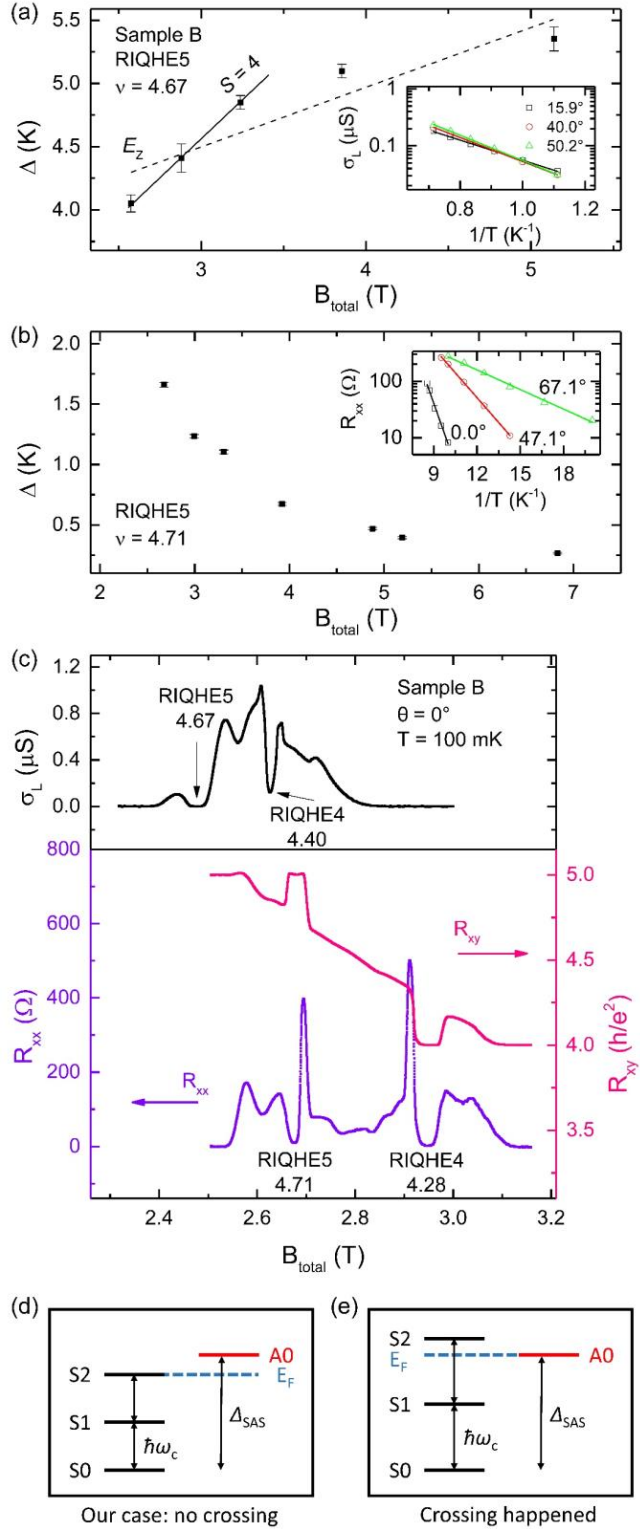

FIG. 4. (a) Anomalous increasing energy gap as a function of total magnetic field is observed at RIQHE5 of Corbino Sample B. The Zeeman energy is determined by $E_Z = g^* \mu_B B_{total} S$. The dashed line is the fitting for effective g-factor with $S = 1$; the solid line is the fitting for $S$ with first 3 points regarding $|g^*| = 0.44$. The result

($S = 4$) shows that the energy enhancement may arise from the existence of skyrmions involving 4 reversed spins. (b) Normal decreasing energy gap as a function of total magnetic field at RIQHE5 of a van der Pauw sample (easy direction) made from the same wafer as Sample B. (c) Longitudinal conductivity of Sample B (upper) and longitudinal and Hall resistance of a van der Pauw sample (easy direction) (lower) made from the same wafer versus total magnetic field at 100 mK without tilt. (d) Schematic for S2 and A0 are not crossed, which is the case of our samples. (e) Schematic for S2 and A0 are crossed. The states in the third Landau level should behave the property of A0.

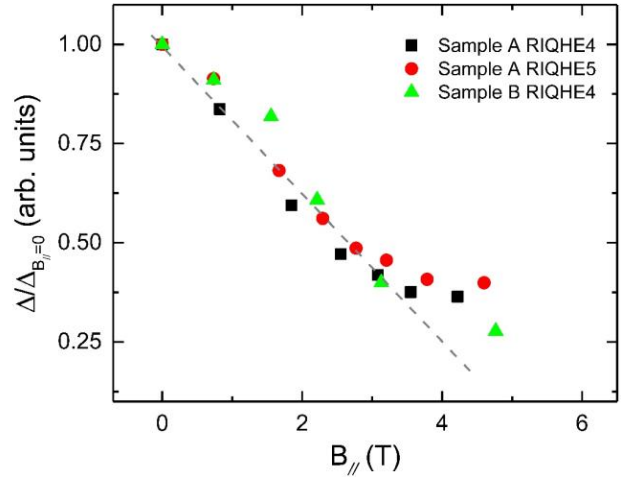

FIG. 5. Normalized energy gap values as a function of in-plane magnetic field are replotted from Fig. 3.